\documentclass[11pt]{article}
\usepackage[margin=1in]{geometry}
\usepackage[T1]{fontenc}
\usepackage{lmodern}
\usepackage{microtype}
\usepackage{graphicx}
\usepackage{booktabs}
\usepackage{amsmath}
\usepackage[hidelinks]{hyperref}
\usepackage{xcolor}

\title{The Long Tail, Not the Front Page:\\
Cold-Start Prediction of Crowd Highlight Salience}
\author{
  Kazuki Nakayashiki \quad Keisuke Watanabe \\[3pt]
  Glasp Inc. \\
  \texttt{kazuki@glasp.co} \quad \texttt{kei@glasp.co} \\[3pt]
  {\small\itshape Co-first authors (equal contribution).}
}
\date{}

\begin{document}
\maketitle

\begin{abstract}
A social highlighter's most useful signal --- which passages a crowd of readers marks ---
exists only for documents people have already read. Can the aggregate crowd salience of a
document be predicted from its text before its marks accumulate? Prior work on this data found
that off-the-shelf zero-shot language models recover highlight locations \emph{worse} than a
trivial lead (position) baseline~\cite{nakayashiki2026selection}, so we ask whether a model
\emph{trained} on the highlight corpus can beat that baseline. Using a pre-registered ladder of
models and a by-document cluster bootstrap, we find a small but robust edge: a logistic ranker
over sentence embeddings and positional/contextual features beats the lead baseline by
$+0.044$ average precision ($95\%$ CI $[+0.029,+0.058]$ from a by-document cluster bootstrap;
clears a pre-registered margin $\delta{=}0.03$ in $97\%$ of resamples, and is stable across
repeated runs of the same dense-document pool). Two closely related unsupervised extractive
baselines (centroid, LexRank-style centrality) \emph{lose} to lead, and the trained model beats
them by $+0.108$, so the edge is not recovered by these generic unsupervised proxies --- it
reflects learning from real reader marks. In product terms the gain is substantial: precision@3
rises from $0.25$ to $0.39$ ($+0.14$ absolute, $+55\%$ relative) and the model has higher
per-document AP than lead on $69\%$ of documents. An ablation shows both the raw embedding
($+0.014$) and training augmentation ($+0.010$) contribute, each with a positive bootstrap CI. The edge is not
a temporal-generalization failure, and we find no evidence that content drift or near-duplicate
leakage explains it (removing near-twins \emph{raises} it). A standardized regression shows the
advantage is governed mainly by document popularity (lower popularity, larger edge) and by label
reliability. It nearly vanishes only on the most popular content; reading both the model's and
the baseline's per-cell scores, that is where the \emph{lead baseline strengthens}, not where
the model weakens. Because our
evaluation conditions on documents that eventually accumulated readers, these results are best
read as a retrospective cold-start simulation; within that scope, day-zero highlight prediction
is most feasible on low-popularity, long-tail documents and least useful on the most popular,
front-page-like content where the lead baseline is already strong.
\end{abstract}

\section{Introduction}
A social web highlighter accumulates, for each document, the spans many readers chose to mark.
Aggregated across readers, these marks are a crowd-salience map and the substrate for
popular-highlight displays, summaries, and discovery. In prior work we showed this aggregate
layer is the \emph{strong} signal in highlighting --- the within-document individual signal is a
whisper (own-versus-other gap $+0.017$ average precision), while what a crowd collectively marks
dominates~\cite{nakayashiki2026salience} --- and that personalizing the salience layer does not
help: an impersonal salience order is not improved by personal re-ranking, and zero-shot
language models predict highlight locations \emph{worse} than a trivial lead
baseline~\cite{nakayashiki2026selection}.

Every application of the aggregate signal shares a structural gap: it needs marks on the
\emph{target} document, and a newly published document has none. We ask whether this gap can be
closed from text: given a document's content, can a model predict where its crowd will
highlight? Because zero-shot prompting is known to fail here~\cite{nakayashiki2026selection},
and because the lead baseline (mark the opening sentences) is a hard bar --- the
recommender-systems analogue, the popularity baseline, is likewise hard to
beat~\cite{krichene2020,revisitpop2020} --- we place the discouraging evidence in front and ask
the falsifiable question: does a \emph{trained} model clear lead by a pre-registered margin?

The findings are: (i) a trained model beats lead by $+0.044$ AP (by-document cluster bootstrap),
and beats two unsupervised extractive baselines that themselves lose to lead; (ii) the gain is
product-meaningful (precision@3 $0.25\to0.39$); (iii) we find no evidence that a temporal
artifact, content drift, or near-duplicate leakage explains it; and (iv) a continuous regression
shows it is governed by document popularity and label reliability, vanishing only on the most
popular content, where the lead baseline --- not the model --- changes. We also document, transparently,
that two intuitive mechanistic stories we initially believed (a temporal shift, then thin
labels as a document property) were each refined by the next control, while the bootstrap-robust
facts persisted.

\paragraph{Contributions.}
(1) To our knowledge the first demonstration, on real reader crowds, that aggregate cold-start
highlight salience is text-predictable beyond the lead baseline by a \emph{trained} model
($+0.044$ AP; precision@3 $+0.14$), in a regime where zero-shot language models and unsupervised
extractive baselines do not beat lead. (2) An ablation attributing the edge to representation
and augmentation, and a robustness battery (temporal fixed-slice, drift, near-duplicate,
multiplicity). (3) A standardized regression and per-cell decomposition localizing the edge to
low-popularity documents and showing the low gap in the most-popular cell is the \emph{baseline} rising, not
the model failing. (4) A methodological note: controls that refined two plausible mechanisms
while the robust facts held.

\section{Related work}
\textbf{Salience and highlights.} Predicting human salience from text is studied in extractive
summarization and highlight detection, usually against reference summaries or small annotator
panels rather than real reader crowds; the behavioral finding that language models model human
salience only weakly~\cite{trienes2025} is the zero-shot ceiling our trained model is measured
against, and reader highlights predict comprehension and interest~\cite{winchell2020}. Nearest are
work on what makes text quotable or memorable~\cite{danescu2012}, automatic pull-quote
selection~\cite{bohn2020}, and industry-facing popular-highlight features on e-reader platforms;
to our knowledge none predicts a reader crowd's aggregate highlight map on cold documents.
\textbf{This program.} Our prior papers locate \emph{where} individuality lives (selection, not
salience)~\cite{nakayashiki2026salience}, show personalization does not help at the salience
layer~\cite{nakayashiki2026selection}, and ask whether the within-document crowd is internally
factional~\cite{nakayashiki2026factions}. Those study the \emph{individual} or the
\emph{segmented} crowd; here we predict the \emph{aggregate} crowd before its marks exist --- a
different object and task. A separate strand studies answer-engine visibility on this
platform~\cite{watanabe2026aeo}.
\textbf{Hard baselines and aggregation.} That a simple baseline (popularity, position) is hard
to beat, and that sampled or mis-specified comparisons distort
conclusions~\cite{krichene2020,revisitpop2020}, motivates our pre-registered lead bar and
bootstrap discipline; the complementary result that aggregating simple predictors can rival a
human crowd~\cite{schoenegger2024,park2024} frames why the aggregate layer is the promising
target.

\section{Data and task}
\textbf{Source and funnel.} We use Glasp, a social web highlighter with hundreds of thousands of
active users and millions of highlighted URLs. From that corpus we sample \emph{dense}
documents (at least $20$ distinct lifetime highlighters) so an aggregate target is stable, fetch
each (live HTTP $+$ readability; yield $\sim$$40$--$60\%$, paywalls/JS/non-HTML excluded),
re-anchor co-reader marks to sentences (excluding social/video domains and capping co-readers at
$60$ per document), and keep documents with at least six anchored co-readers. Each evaluation
run yields $\sim$$284$ usable evaluation documents and $\sim$$389$ training-only documents
(Table~\ref{tab:data}). Because we condition on documents that \emph{eventually} reached $20$
highlighters, the evaluation population is a survivorship-filtered slice, not a true day-zero
population; we treat ``cold start'' as a retrospective simulation and return to this in
Limitations.

\begin{table}[t]
\centering
\caption{Dataset summary for one evaluation run.}
\label{tab:data}
\begin{tabular}{lc}
\toprule
Candidate dense documents sampled ($\geq 20$ highlighters) & $705$ \\
\quad $\to$ evaluation documents (after fetch $+$ anchoring gate) & $284$ \\
Training-only candidates ($\in[12,20)$) $\to$ usable & $1{,}045 \to 389$ \\
Sentences (evaluation) & $32{,}549$ \\
Median sentences / document & $72$ (range $8$--$400$) \\
Median anchored co-readers / document & $21$ (range $6$--$60$) \\
Median lifetime highlighters / document & $27$ (range $20$--$592$) \\
Positive-class base rate (top $15\%$) & $0.146$ \\
Article fetch yield & $\sim$$40$--$60\%$ \\
\bottomrule
\end{tabular}
\end{table}

\textbf{Anchoring and label.} Glasp does not store article bodies, so we fetch and split into
sentences and re-anchor each co-reader's stored highlight strings to sentence indices by text
matching (substring, then token-Jaccard $\geq 0.6$), keeping a reader only if at least half of
their spans anchor, following our prior work~\cite{nakayashiki2026salience}. Each sentence's
crowd count is the number of anchored co-readers; the target is the top $15\%$ of sentences by
count (base rate $\approx 0.146$; ties in crowd count broken by a fixed per-sentence jitter, so
the threshold is well defined even at low reader counts). All models and baselines rank the
\emph{same} labels.

\textbf{Metric and inference.} We report average precision (AP) per document and the
\emph{advantage} of a model over lead, $\mathrm{AP}(\text{model})-\mathrm{AP}(\text{lead})$, with
a $3{,}000$-iteration cluster bootstrap resampling whole documents; ``frac${>}0$'' and
``frac${>}\delta$'' are the fractions of resamples with advantage above $0$ and above the
pre-registered margin $\delta{=}0.03$. For products we also report precision@$k$.
\emph{Pre-registered} before running: the model ladder, $\delta$, the pass rule, and the primary
$\mathrm{AP}(\text{model})-\mathrm{AP}(\text{lead})$ comparison; the pass rule is a point estimate
$\geq \delta$ \emph{and} every by-document resample positive (frac${>}0=1$). The ablation,
unsupervised baselines, precision@$k$, and the regression are planned robustness and
interpretability analyses added after the primary result. (Random AP is $0.186$, slightly above
the $0.146$ base rate: on short documents the expected AP of a random ranking exceeds the
positive rate, a finite-list effect, not a bug.) The pipeline runs on private user data and is
not released.

\section{Models}
A pre-registered ladder of class-balanced logistic rankers over sentences:
\textbf{M0}, position $+$ lexical features (normalized position, a lead indicator, log length,
TF-IDF); \textbf{M1}, a raw \texttt{text-embedding-3-large} sentence embedding (requested at
$768$ dimensions via the API \texttt{dimensions} parameter) $+$ position; \textbf{M2}, M0 $+$ derived semantic/contextual features (cosine to the document
centroid, degree centrality, cosine to the lead, cosine to neighbors, a novelty score, surface
cues); \textbf{M3}, the combined model (raw embedding $+$ the M2 feature set) trained on the
dense pool \emph{augmented} with training-only documents of moderate popularity
($\textit{userCount}\in[12,20)$), which enlarges training without changing the evaluation
population. A one-hidden-layer MLP on the M2 features tests nonlinearity (it does not help).
Baselines: \textbf{random}; \textbf{lead} (earlier sentences ranked higher), the bar to clear;
and two closely related \emph{unsupervised extractive} baselines, \textbf{centroid} (cosine to
the document embedding centroid) and \textbf{centrality} (mean cosine of a sentence to every
other sentence, a
degree-centrality / LexRank-style score; on $L_2$-normalized embeddings it nearly coincides with
the centroid baseline, hence their near-identical scores). M0 underperforming lead is expected: a
trained ranker that must generalize across documents does not beat the raw within-document
position order until richer features are added.

\section{Results}

\subsection{An edge over lead, attributed, and product-meaningful}
\begin{figure}[t]
\centering
\includegraphics[width=0.74\linewidth]{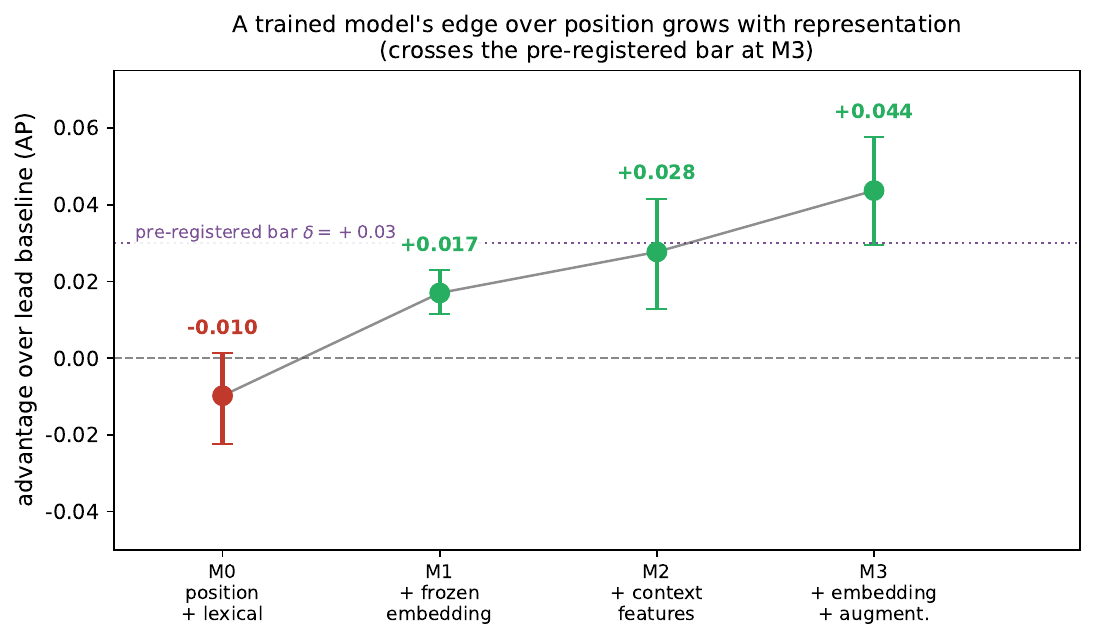}
\caption{The trained model's edge over lead grows with representation and crosses the
pre-registered bar ($\delta{=}0.03$) at M3 (staged pre-registration runs; Table~\ref{tab:ablation}
gives the single-run ablation). Points are mean advantage; bars are $95\%$ by-document CIs.}
\label{fig:ladder}
\end{figure}

The advantage over lead grows with representation and the headline model M3 beats lead by
$+0.044$ AP (Table~\ref{tab:ablation}). It meets the pre-registered pass rule (point estimate
$+0.044 \geq \delta$; frac${>}0=1.0$); as a stricter check it exceeds $\delta$ in $97\%$ of
resamples, and the de-duplicated estimate (\S5.4) excludes $\delta$ at the CI level. (The margin
was a pre-registered decision threshold for the \emph{point estimate}, not a CI-exclusion
criterion; we report frac${>}\delta$ as a stricter descriptive check.)
Figure~\ref{fig:ladder} reports the staged pre-registration runs (M0--M3); Table~\ref{tab:ablation}
is a separate run of the same small pool used for the ablation (the dense-document pool is small
and the runs overlap heavily; \S5.4), so M2 differs slightly ($+0.028$ vs $+0.024$) across runs.
The ablation attributes the M2$\to$M3 gain to both knobs: the raw embedding contributes
$+0.014\,[+0.010,+0.019]$ and the augmentation $+0.010\,[+0.002,+0.018]$, each with a positive
bootstrap CI. (The last-mile contributions are weakly superadditive --- adding each knob alone to
M2 gives $+0.010$ and $+0.005$ --- so these values include the interaction and slightly overcount
the separate effects.) Crucially, the two unsupervised extractive baselines lose to lead
($-0.065$), and M3 beats centrality by $+0.108$: the edge is not recovered by these generic
unsupervised proxies --- it comes from learning real reader marks (a supervised baseline trained
on summaries is untested; see Limitations). In product terms the gain is large: precision@3
rises from $0.254$ to $0.394$ and precision@5 from $0.259$ to $0.346$ (Table~\ref{tab:product}),
and M3 has higher per-document AP than lead on $69\%$ of documents (ties counted as non-wins).

\begin{table}[t]
\centering
\caption{Ablation and baselines (the ablation run: $284$ documents, a separate run of the same
pool as Runs A--C; advantage $= \mathrm{AP}(\cdot)-\mathrm{AP}(\text{lead})$, $95\%$ by-document
cluster-bootstrap CI). Unsupervised extractive baselines lose to lead; both M3 knobs contribute.}
\label{tab:ablation}
\begin{tabular}{llcc}
\toprule
Method & Note & Advantage over lead & frac${>}\delta$ \\
\midrule
centroid (unsup.)      & cosine to doc centroid     & $-0.065\,[-0.092,-0.038]$ & $0.00$ \\
centrality (unsup.)    & LexRank-style              & $-0.065\,[-0.093,-0.039]$ & $0.00$ \\
M2                     & context features           & $+0.024\,[+0.009,+0.040]$ & $0.24$ \\
M2 $+$ augmentation    &                            & $+0.029\,[+0.015,+0.043]$ & $0.43$ \\
M2 $+$ embedding       &                            & $+0.034\,[+0.018,+0.049]$ & $0.68$ \\
\textbf{M3} & \textbf{$+$ embedding $+$ augmentation} & $\mathbf{+0.044\,[+0.030,+0.058]}$ & $\mathbf{0.97}$ \\
\midrule
\multicolumn{2}{l}{embedding effect (M3 $-$ M2$+$aug)} & $+0.014\,[+0.010,+0.019]$ & --- \\
\multicolumn{2}{l}{augmentation effect (M3 $-$ M2$+$emb)} & $+0.010\,[+0.002,+0.018]$ & --- \\
\bottomrule
\end{tabular}
\end{table}

\begin{table}[t]
\centering
\caption{Product-facing metrics (top-$k$ precision; same run). The model substantially improves
top-$k$ precision (a $+55\%$ relative gain at $k{=}3$).}
\label{tab:product}
\begin{tabular}{lccc}
\toprule
Metric & lead & M3 & M3 $-$ lead \\
\midrule
precision@3 & $0.254$ & $0.394$ & $+0.141\,[+0.103,+0.180]$ \\
precision@5 & $0.259$ & $0.346$ & $+0.087\,[+0.059,+0.116]$ \\
\bottomrule
\end{tabular}
\end{table}

\subsection{Not a temporal-generalization failure}
Testing on the newest slice of documents, the advantage falls to $+0.009$ (not significant). But
this is \emph{test-set composition}, not train/test time order. Fixing a future test slice $T$
and scoring the same $T$ two ways --- trained on the \emph{past} only versus a same-era
cross-validation within $T$ --- gives statistically identical results (contemporaneous $+0.012$
vs temporal $+0.011$; difference $\Delta +0.001\,[-0.008,+0.010]$). (The earlier $+0.009$ and this
$+0.011$ differ only because they use different test sets: the newest cross-validation fold versus
the separately defined fixed future slice $T$.) There is no temporal-generalization failure; the
newest slice is simply a harder \emph{population} (\S5.3). We initially read the $+0.009$ as a
temporal shift; the fixed-slice control retracts that.

\subsection{Where the edge lives: popularity, and the baseline, not the model}
\begin{figure}[t]
\centering
\includegraphics[width=\linewidth]{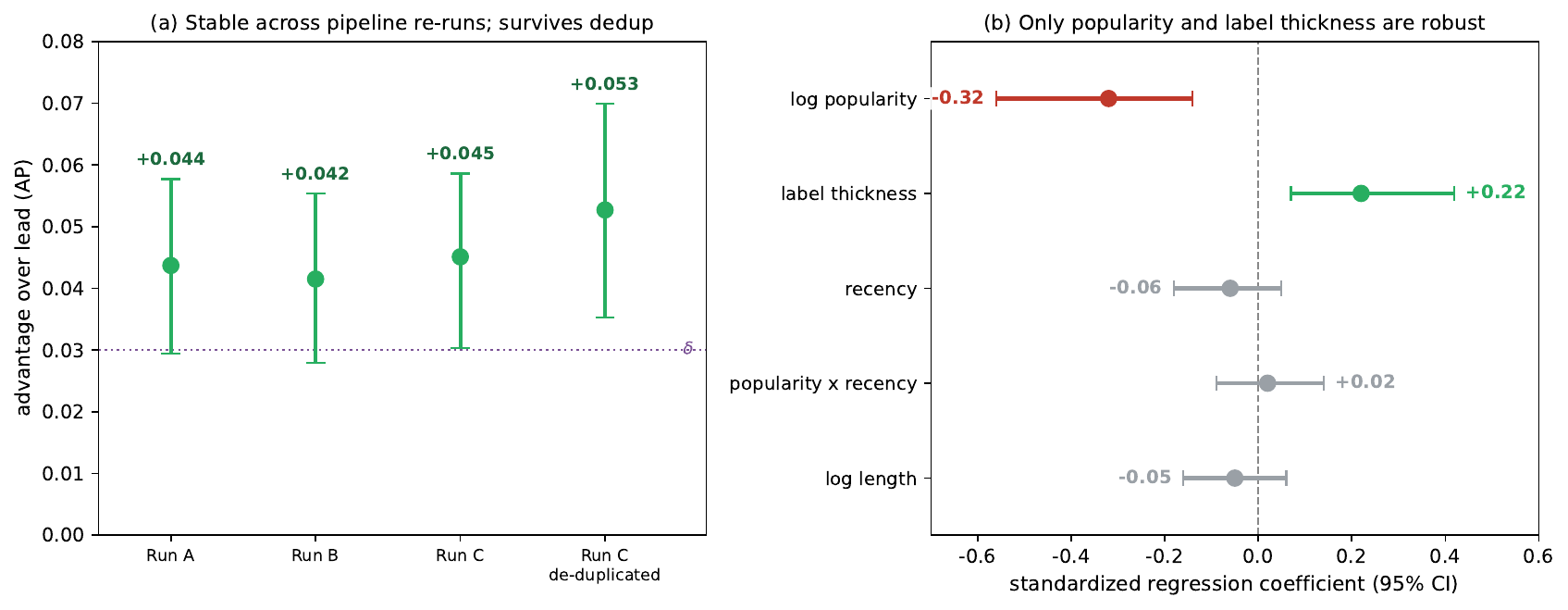}
\caption{(a) The advantage over lead is the same across three pipeline re-runs of the small
document pool (Runs A--C, $\sim$$98\%$ overlapping) and rises after removing near-duplicates;
every bootstrap positive. (b) Standardized
regression of the per-document advantage: only document popularity ($-0.32$) and label thickness
($+0.22$) are robust; recency, the popularity$\times$recency interaction, and length are not.}
\label{fig:robust}
\end{figure}

A standardized regression of the per-document advantage on document
covariates\footnote{OLS of $\mathrm{AP}(\text{M3})-\mathrm{AP}(\text{lead})$ on standardized log
popularity, recency (the document's median co-reader highlight timestamp), their product, log
label thickness (anchored co-readers), and log length; $95\%$ by-document bootstrap percentile
CIs. Popularity and label thickness correlate $r=0.69$.} (Figure~\ref{fig:robust}b) shows two
robust drivers: lower document popularity ($\beta=-0.32\,[-0.56,-0.14]$) and thicker crowd labels
($\beta=+0.22\,[+0.07,+0.42]$). Recency, a popularity-by-recency interaction, and document length
are \emph{not} significant once controlled --- so the heterogeneity is a popularity \emph{main
effect}, not the interaction a simpler analysis suggested. This reconciles with the per-cell
pattern: the high-popularity-and-recent cell concentrates the \emph{most} popular documents (mean
lifetime highlighters $81$, versus $55$ in the high-popularity-established cell), so the
continuous popularity main effect alone accounts for its low advantage --- no interaction term is
needed.

\begin{figure}[t]
\centering
\includegraphics[width=0.86\linewidth]{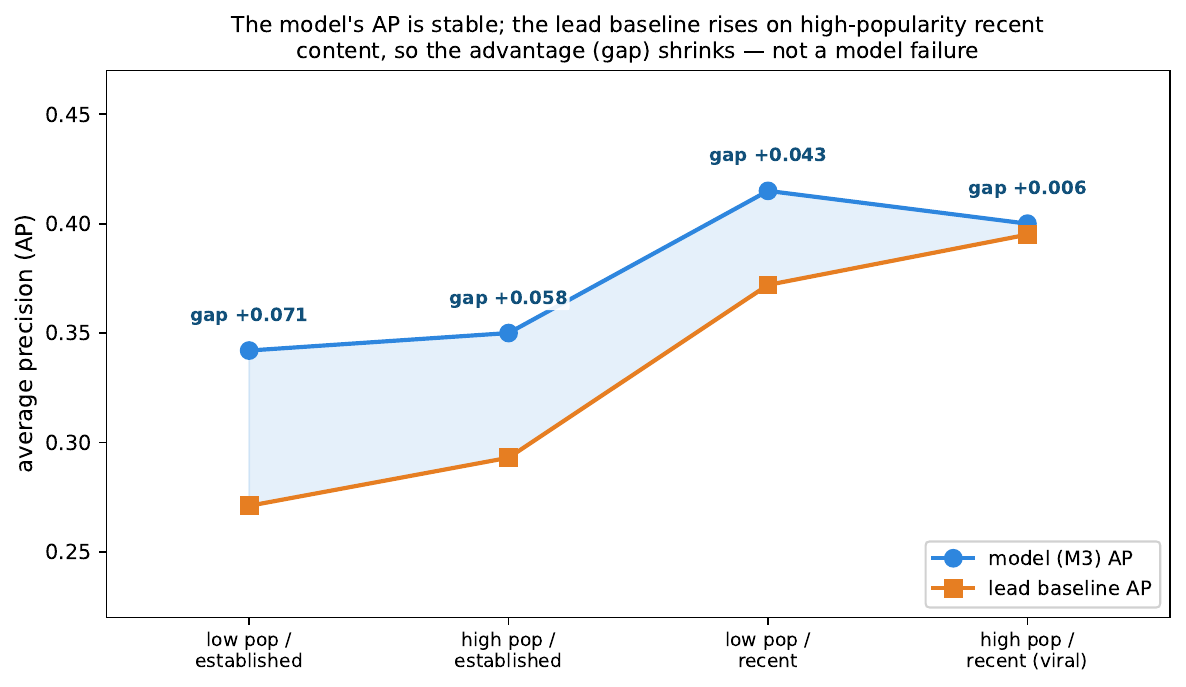}
\caption{Per-cell model AP and lead AP across popularity $\times$ recency (median splits). The
model's AP is stable; the \emph{lead} AP rises on the high-popularity, recent cell (which
concentrates the most popular documents, mean $81$ lifetime highlighters), which is why the
advantage (the gap) shrinks there. The edge is largest where lead is weak, not where the model
fails.}
\label{fig:grid}
\end{figure}

Reading both the model's and the baseline's scores per cell is decisive (Figure~\ref{fig:grid}).
The model's absolute AP is roughly stable across the popularity$\times$recency grid
($0.34$--$0.42$); what varies is the \emph{lead} AP, which jumps from $\sim$$0.29$ on established
documents to $0.39$ on the most popular, recent ones. So the advantage nearly vanishes there not
because the model weakens but because \emph{position becomes nearly as good}: on the most popular
content the crowd concentrates on the opening. The honest statement is ``the edge is largest
where the lead baseline is weak.''

Finally, the label-thickness effect at least partly reflects \emph{measurement reliability}:
documents with thin crowd labels (few anchored co-readers, a noisier target) show no significant
edge ($+0.019\,[-0.010,+0.045]$), while mid and thick labels show $+0.05$--$0.06$. A noisier
target attenuates any model$-$lead gap, so $+0.044$ could be conservative; we caution, however,
that thin-label documents are also the closest proxy to a true zero-reader document, so whether
the genuinely-new regime is harder to \emph{predict} or merely harder to \emph{measure} remains
an open uncertainty rather than a settled lower bound.

\subsection{Robustness: drift, near-duplicates, anchoring, multiplicity}
\textbf{Content drift.} We fetch the current page, whereas readers marked a possibly earlier
version. We find no evidence that drift explains the effect: the cleanest (most-recent,
least-drifted) documents are not the strongest, and the popularity effect survives controlling
recency. We cannot fully exclude drift without archived snapshots, which we leave to future work.
\textbf{Near-duplicate leakage.} Scoring each test document's maximum embedding-centroid cosine
to any training document, only $5\%$ have a $\geq 0.95$ near-twin, and removing them
\emph{raises} the advantage (overall $+0.045\to$ clean $+0.047\to$ clean-strict $+0.053$;
Table~\ref{tab:robust}); the headline is conservative on a deduped corpus.
\textbf{Anchoring noise.} Because all methods rank the same labels, anchoring noise cannot by
itself create method-specific labels; however, feature-correlated anchoring (e.g.\ if lead or
boilerplate sentences anchor differently) remains a limitation we cannot fully rule out.
\textbf{Multiplicity.} We computed many advantage contrasts; we therefore lead with the
by-document bootstrap CI on the evaluation set (every resample positive) and its stability across
pipeline re-runs, rather than the maximum over candidates.

\begin{table}[t]
\centering
\caption{Robustness of the headline advantage. The dense-document pool is small ($\sim$$288$
usable documents); the three runs (A--C) each sample nearly all of it and overlap $\sim$$98\%$,
so they are essentially one evaluation set and their agreement is a consistency check across
pipeline re-runs, not independent replication. The primary inference is the by-document bootstrap
CI on those documents. De-duplication raises the edge; the temporal control shows no shift.}
\label{tab:robust}
\begin{tabular}{llc}
\toprule
Check & Run / contrast & Advantage over lead \\
\midrule
Stability & Run A (primary) & $+0.044\,[+0.029,+0.058]$ \\
Stability & Run B & $+0.042\,[+0.028,+0.055]$ \\
Stability & Run C & $+0.045\,[+0.030,+0.059]$ \\
Near-dup  & Run C, near-twins removed & $+0.053\,[+0.035,+0.070]$ \\
Temporal  & contemporaneous $-$ past-only (same slice) & $+0.001\,[-0.008,+0.010]$ \\
\bottomrule
\end{tabular}
\end{table}

\subsection{A methodological note}
Two intuitive explanations we first believed were refined by the next control. We read the weak
newest-slice result as a temporal-generalization failure; the fixed-slice control (\S5.2) showed
past-only and same-era training are identical. We then read the weakness as thin/young labels as
a document property; the regression and per-cell scores (\S5.3) showed label thickness is a
\emph{reliability} moderator and that the viral cell reflects a rising baseline. Throughout, the
bootstrap-robust facts --- the $+0.044$ edge, its stability across re-runs, the absence of a temporal shift,
the unsupervised baselines losing to lead --- did not move. We recommend leading with
resampling-robust effect sizes and treating each mechanistic ``why'' as a hypothesis the next
control must try to kill.

\section{Discussion}
\textbf{Relation to the program.} Personal Salience showed the crowd dominates the individual
within a document~\cite{nakayashiki2026salience}; Selection, Not Salience showed personalization
does not help at the salience layer and that aggregating beats
personalizing~\cite{nakayashiki2026selection}. This paper predicts the aggregate layer those
papers identified as strong, on documents lacking marks. There is no contradiction with the
$+0.017$ individual ``whisper'': that measured a person's history predicting \emph{their own}
marks, whereas here the crowd is the \emph{target}.
\textbf{Product reading.} The cold-start use case --- pre-fill popular highlights on a new
document --- is exactly the low-popularity region, where the edge is positive and bootstrap
robust and where organic crowd signal is likely to be slower or sparser. On the most popular
content the model and lead are close because position already works and a human crowd arrives
fast, so prediction is least needed precisely where it is least distinctive. A day-zero feature
is thus most useful on the long tail, with the honest caveat that our evaluation conditions on
documents that did eventually reach a crowd.

\section{Limitations}
\textbf{Survivorship / simulated cold start.} We evaluate on documents that eventually reached
$\geq 20$ highlighters; this is not a true zero-reader population, and the genuinely-new regime
is approximated, not measured. \textbf{Label reliability.} Thin labels attenuate the measured
edge (\S5.3), so $+0.044$ may be conservative; but the thinnest, most cold-start-like documents
are also the noisiest, so whether the genuinely-new regime is harder to \emph{predict} or merely
harder to \emph{measure} is unresolved. \textbf{Population skew.} Fetchable dense documents skew
short and popular; long/niche content and the $\sim$$40$--$60\%$ fetch loss are survivorship
filters. \textbf{Anchoring and drift.} Live fetching and text anchoring introduce noise we
control for but cannot eliminate; an archived-snapshot replication and a domain-held-out split are
natural next steps. \textbf{Label definition.} The top-$15\%$ binary target is one choice; a
count-weighted target is a natural robustness extension. \textbf{Model ceiling.} We test logistic
rankers over frozen embeddings; a fine-tuned encoder and a supervised baseline trained on
summaries are untested, so M3 is not an upper bound (and the unsupervised result bounds only
\emph{unsupervised} alternatives). \textbf{Magnitude.} The edge is real and robust but modest; it
should be read as ``usefully better than position on most documents,'' supported by the
precision@3 gain, not as near-oracle prediction.

\section{Ethics}
This study analyzes highlighting behavior on a consumer platform under its terms of service. It
is internal, aggregate-level analytics rather than experimental intervention: no individual is
profiled, no individual-level claims are reported, and no model is trained to predict any
individual user --- user identifiers are used
only to form document-level crowd counts, and documents below a minimum aggregation threshold
(fewer than six anchored co-readers) are excluded. We report no individual user's marks, release
no per-user or per-document behavioral data, and the extraction/scoring pipeline is not published;
highlighting patterns can be identifying, so only bootstrap estimators and aggregate statistics
are shared, on reasonable request, and deletion/private-content requests are honored upstream.
Predicted highlights are an editorial convenience and carry the usual risk of nudging attention;
the popularity gradient we report (weakest on the most popular material) limits the mechanism's
reach over the highest-traffic content.

\section{Conclusion}
On real reader crowds, the aggregate salience of a document is text-predictable beyond the lead
baseline by a trained model ($+0.044$ AP; precision@3 $0.25\to0.39$), in a regime where zero-shot
language models and unsupervised extractive baselines do not beat lead. We find no evidence that
a temporal artifact, content drift, or near-duplicate leakage explains the edge; it is governed
by document popularity and label reliability, and it shrinks on the most popular content because
the lead baseline strengthens there, not because the model fails. Within the scope of a
retrospective cold-start simulation, the result places day-zero highlight prediction where it is
most useful --- the long tail --- and the
journey reframes a recurring hazard: resampling-robust effect sizes outlive the mechanistic
stories attached to them.

\section*{Reproducibility}
Every effect size carries a $3{,}000$-iteration by-document cluster-bootstrap CI; the model
ladder, the margin $\delta$, and the slice definitions were fixed before running. The
data-extraction and scoring pipeline runs against Glasp's private user data and is \textbf{not}
released; per-document results derive from individual highlighting behavior and are not published.
The cluster-bootstrap estimator and aggregate statistics are available to researchers on
reasonable request.

\end{document}